\documentstyle[11pt]{article}
\begin{document}
\newcommand{\DUR}{\em Department of Physics\\
                      University of Durham \\
                      Science Laboratories, South Road\\
                      Durham, DH1 3LE, U.K.}
\newcommand{\NPB}[3]{{\em Nucl.Phys.} {\bf B#1} (19{#2}) {#3}}
\newcommand{\PLB}[3]{{\em Phys.Lett.} {\bf B#1} (19{#2}) {#3}}
\newcommand{\PRD}[3]{{\em Phys.Rev.}  {\bf D#1} (19{#2}) {#3}}
\newcommand{\PRL}[3]{{\em Phys.Rev.Lett.} {\bf #1}(19{#2}){#3}}
\newcommand{\IJA}[3]{{\em Int.J.Mod.Phys.} {\bf A#1} (19{#2}) {#3}}
\def\QQa{\renewcommand{\baselinestretch}{1.3}\Huge\large\normalsize}

\def\bx{{\bf x}}
\def\bp{{\bf p}}
\def\pa{\partial} \def\da{\dagger} \def\dda{\ddagger}
\def\partialslash{\partial\!\!\!/}
\def\pslash{p\!\!\!/}
\def\lhs{{\it l.h.s.}}
\def\rhs{{\it r.h.s.\ } }
\def\ph4{\lambda\phi{4}} \def\bfnc{$\beta$-function}
\newcommand{\dr}{differential renormalization }

\def\bqn{\begin{equation}}
\def\eqn{\end{equation}}
\def\bqna{\begin{eqnarray}}
\def\eqna{\end{eqnarray}}
\def\bit{\bibitem}
\def\vs{\vspace{.25in}}
\def\thefootnote{\fnsymbol{footnote}}

\pagestyle{empty}
{\hfill \parbox{6cm}{\begin{center} DTP/92/34 \\
                                    May 1992
                     \end{center}}}
\vspace{1.5cm}

\begin{center}
\large{\bf  A NEW APPROACH TO\\
          LOWER DIMENSIONAL GAUGE THEORIES}
\vskip .6truein
\centerline {Ramon Mu\^noz-Tapia\footnote{e-mail: RMT@HEP.DUR.AC.UK}}
\end{center}
\vspace{.3cm}
\begin{center}
\DUR
\end{center}
\vspace{1.5cm}

\centerline{\bf Abstract}
\medskip

We apply the method of differential renormalization to two and three
dimensional abelian gauge theories.
The method is especially well suited for these theories as the
problems of defining the antisymmetric tensor are avoided
and the calculus involved is impressively simple.
The topological and dynamical photon masses are obtained.

\newpage
\pagestyle{plain}
\QQa

In a recent paper [1] a new method of renormalization has been introduced.
The differential renormalization procedure consists of three steps:
$i)$ write the amplitudes in coordinate space,
$ii)$ locate the factors that are too singular at
short distances to allow for a Fourier transform, and
$iii)$ write these factors
as derivatives of less singular expressions. The derivatives have to be
understood as integrated by parts, {\it i.e.} surface terms are dropped.
The approach gives finite expressions corresponding to the renormalised
amplitudes in a certain scheme. The easiest example is illustrated in massless
$\ph4$ theory, where the one loop four-point amplitude contains the factor
$1/x4$, which is regulated by the differential equality
\bqn
{1\over x4} = -{1\over 4} \Box{\ln (x2M2)\over x2} \ \ .
\label{i}
\eqn

\noindent The above expression is an identity as long as $x \neq 0$. The
{\it l.h.s.\ }is too singular at $x=0$, but the {\it r.h.s.\ },
due to the presence of the
differential operator, can be Fourier transformed once the surface terms
are ignored. The parameter $M$ is an integration constant that plays the
role of a renormalization scale.

Differential renormalization has been shown to be very powerful in a three
loop computation of $\ph4$
and has already been applied in different contexts [1,2,3]. Although a general
proof of the consistency of the method is missing, in Ref.\ [4] a computation
at the three loop level has been redone with an explicit cutoff. The divergent
terms neglected in the partial integration are made explicit and it is seen
that they are indeed cancelled by counterterms of the wave function, mass, and
coupling constant redefinition. Very recently compact expressions similar
to Eq.\ (1) have been obtained for massive cases [5].

The purpose of the present note is to show that the method of \dr is
specially suited for massive or massless three and two dimensional theories.
These theories exhibit many interesting features related to the appearance of
topological [6,7] and dynamical symmetry  breaking effects [8]. There is
also the hope that they can play an important role in the understanding of
high $T_c$ superconductivity and the Quantum Hall Effect [9,10]. It is known
that in QED(2+1) the photon acquires a mass of topological origin if
massive fermions are present. Coleman and Hill [11] proved that only
the one loop correction contributes to it. The use of dimensional
regularization here is rather cumbersome due to the presence of the
antisymmetric tensor $\epsilon_{ijk}$ and the problems of extending its
definition to arbitrary dimensions.
One of the nice features of \dr is that
space-time remains untouched, thus avoiding problems associated with
defining quantities in arbitrary dimensions.
The regularization is achieved by keeping
the coincident points apart by means of equalities such as (1)
or similar.

We  first analyse  massless three dimensional QED by computing the photon
and fermion self energy. The results agree with Ref.\ [12]. Next we deal with
the massive case, where the one loop correction to the photon self energy
induces a topological mass which is computed. Finally we consider the two
dimensional case, where the photon mass is induced by the so called
Schwinger mechanism [8]. The results are shown to be gauge invariant in a
very transparent manner, to be compared with dimensional or other
regularization methods [13], where the gauge breaking parts cancel in a
complicated way.

We will work in Euclidean space, where the Lagrangean reads

\bqn
L(x)= \frac14 F_{ij}F_{ij} +
     \bar{\Psi} (\partialslash + i \, e {A \!\!\!/} + i m )\Psi
\label{ii}
\eqn
with
\bqn
F_{ij}=\partial_i A_j - \partial_j A_i \ \ .
\eqn

\noindent The Euclidean three dimensional algebra can be realised as [6]
\bqn
\gamma_i=i \sigma_i \ \ ,
\eqn
where $\sigma_i$ are the $2\times 2$ Pauli matrices. The $\gamma_i$ have
the property
\bqn
\gamma_i \gamma_j = -\delta_{ij} - \epsilon_{ijk} \gamma_k \ \ .
\eqn

\noindent The propagator for the  fermion and the boson in coordinate space
are,
respectively,
\bqn
S(\bx)=(\partialslash - im) \Delta(x)
\eqn
and
\bqn
\Delta_{ij}(\bx)=\delta_{ij} \Delta(x) \ \ ,
\eqn
this latter written in the Feynman gauge and $\Delta(x)$ given by
\bqn
\Delta(x)=\frac{1}{4\pi} \frac{\mbox{e}{-m x}}{x} \ \ .
\eqn
The one loop vacuum polarization is
\bqn
\Pi_{ij}(\bx)= -(-ie)2 \mbox{Tr}\left( \gamma_i S(\bx) \gamma_j
              S(-\bx) \right) \ \ .
\eqn
In the massless case it reduces to
\bqn
\Pi_{ij}(\bx)=- \frac{e2}{8 \pi2} \left( 2 \frac{x_i x_j}{x2}
                    - \delta_{ij} \right) \frac{1}{x4} \ \ ,
\eqn
where the following trace properties of the matrices in Eq. (4) have been
used:
\bqn
\mbox{Tr}(\gamma_i \gamma_j)= -2\delta_{ij} \ \ \ , \ \ \
\mbox{Tr}(\gamma_i \gamma_j \gamma_k)= 2 \epsilon_{ijk} \ \ .
\eqn

The structure of (10) suggests that $\Pi_{ij}$ can be written as a two
differential operator acting on a scalar function.
Dimensional analysis shows that the function can
only be $1/x2$, thus
\bqn
\Pi_{ij}(\bx)=-\frac{e2}{32 \pi2} (\partial_i \partial_j - \Box
\delta_{ij}) \frac{1}{x2} \ \ .
\eqn
The last factor has a well defined Fourier transform, the differential
operator in front has reduced the degree of divergence in two units and no
further differential equalities are needed. Hence, no scale similar to $M$
appears. Notice that (12) is automatically gauge invariant
\bqn
\partial_i \Pi_{ij}(\bx)=0 \ \ .
\eqn

\noindent To obtain the momentum space version of Eq.\ (12) the differential
operator  has to be integrated by parts,
while the remaining factor has a trivial Fourier transform proportional to
$1/p$. We have then
\bqn
\Pi_{ij}(\bp)=-\frac{e2}{16} p \left( \delta_{ij}-\frac{p_i
p_j}{p2} \right) \ \ ,
\eqn
which agrees with Ref.\ [12]. The differential equality leading from (10) to
(12) has successfully coped with the linear divergence that would have
appeared had we worked directly in momentum space.

In a similar fashion one can compute the fermion self energy. In coordinate
space the self energy is
\bqn
\Sigma(\bx)=-e2 \gamma_i S(\bx)\gamma_j\Delta_{ij}(\bx)
\ \ .
\eqn
Using the definitions (6) and (7) one obtains in the massless case
\bqn
\Sigma(\bx)=-\frac{e2}{32 \pi2} \partialslash
\left(\frac{1}{x2}\right)
\eqn
This expression is linear in the gauge parameter and vanishes in the
Landau gauge. In momentum space this becomes
\bqn
\Sigma(\bp)=i\frac{e2}{16}\frac{\pslash}{p}
\eqn

Let us now turn our attention to the more interesting massive case. Using
(11) to perform the trace factors, one obtains the following expression
for the polarization tensor
\bqna
\Pi_{ij}(\bx)&=&-2 e2 \left( 2\partial_i \Delta (x) \partial_j \Delta (x)-
\delta_{ij} \left( \partial_k
\Delta(x)\partial_k\Delta(x)+m2\Delta2(x)\right) \right) \cr
&&+4 i e2 m \epsilon_{ijk} \Delta(x) \partial_k \Delta (x)
\eqna
This expression can be put into a form analogous to Eq. (12) with the extra
antisymmetric factor
\bqn
\Pi_{ij}(\bx)=-e2 (\partial_i \partial_j - \Box
\delta_{ij}) F(x) + im e2 \epsilon_{ijk}\partial_k G(x) \ \ .
\eqn
Using the property of $\Delta(x)$
\bqn
\Delta{\prime} (x)=-\Delta(x)\left( m+\frac{1}{x} \right) \ \ ,
\eqn
where the prime denotes derivative of the modulus $x$, one can see that $F(x)$
satisfies the following differential equations
\bqna
F{\prime\prime} (x) - \frac{F{\prime} (x)}{x} &=&
4 \Delta2(x)\left( m+\frac{1}{x} \right)2 \nonumber \\[1ex]
F{\prime\prime} (x) + \frac{F{\prime} (x)}{x} &=&
2 \Delta2(x)\left( 2m2+\frac{2m}{x} + \frac{1}{x2} \right)
\eqna
Equations (21) can be solved, yielding a compatible solution given by
\bqn
F(x)=\frac{1}{32 \pi2}\left( \frac{\mbox{e}{-2mx}}{x2}+ 2 \frac{m
\mbox{e}{-mx}}{x}+4 m2 \mbox{Ei}(-2mx) \right) \ \ ,
\eqn
where Ei$(x)$ is the exponential-integral function [14]. $F(x)$ has a well
defined Fourier transform, $F(p)$, denoted by
\bqn
e2 p2 F(p)=\Pi{(1)}(p2) \ \ .
\eqn
Notice that $\Pi{(1)}(0)=0$, as gauge invariance dictates for the
transverse part, and that in the limiting case $m\rightarrow 0$,
the result (14) is trivially recovered.

As it is shown below, the last term of Eq.\ (18) is the one responsible for
the topological photon mass induced at the one loop level. Since it is
proportional to $m$, it is only present in the massive version of (2).
It is clear that this has to be so, as
massive fermion terms in the realization (3) violate parity [6]. A
parity-violating, but gauge invariant, mass term for the photon can then appear
due to quantum corrections. The function $G(x)$ is easily found to be
\bqn
G(x)=\frac{1}{8\pi2}\frac{\mbox{e}{-2mx}}{x2} \ \ .
\eqn
Its Fourier transform, $G(p)$, is
\bqn
-e2 G(p)=\Pi{(2)}(p2)=\frac{-e2}{2\pi p}
          \arctan\left(\frac{p}{2m}\right) \ \ .
\eqn

Now we can see the effect of the polarization tensor in the complete photon
propagator $D_{\mu\nu}$. In the more familiar Minkowski space it reads [6]
\bqn
D_{\mu\nu}(p)=\frac{-i}{p2-\Pi(-p2)}\left(
g_{\mu\nu}-\frac{p_{\mu} p_{\nu}}{p2}-i \epsilon_{\mu\nu\alpha}
\frac{p\alpha}{p2} M(-p2) \right) \ \ ,
\eqn
with
\bqn
\Pi(p2)=\Pi{(1)}(p2)+m\Pi{(2)}(p2)M(p2) \ \ ,
\eqn
and
\bqn
M(p2)=\frac{m \Pi{(2)}(p2)}{1-\Pi{(1)}(p2)} \ \ ,
\eqn
where $\Pi{(1)}$, $\Pi{(2)}$ are defined in Eqs.\ (23), (25) and
p is the Minkowski momentum.

\noindent Since $\Pi{(1)}(0)=0$, a topological mass is only generated if
$\Pi{(2)}(0) \neq 0$. From Eq.\ (25) one readily sees
\bqn
\Pi{(2)}(0)=\frac{-e2}{4 \pi m} \ \ .
\eqn
At the action level, $\Pi{(2)}(0)$ is the factor that appears in
front of the induced Chern-Simons term $\epsilon_{\mu\nu\rho} F{\mu\nu}
A{\rho}$ [15].
The result (29) agrees with other regularization methods [13,16],
but not with Pauli-Villars, which is a questionable method in
this environment [6].

We finally consider the 2-dimensional case or Schwinger model
[8]. Here a photon mass is also generated, but now by another
mechanism, namely $\Pi{(1)}(0)\neq 0$. Unlike the previous example, the
interesting symmetry breaking effects are already present in the massless
version, which  for simplicity is the only one we consider.
The equivalent of Eq. (7) in the
massless case is
\bqn
\Delta(x)=-\frac{1}{2\pi}\ln(a x) \ \ ,
\eqn
where $a$ is an irrelevant  constant needed for dimensional reasons.
An analysis similar
to Eqs. (9-12) with the standard realization of the Dirac algebra leads to
\bqn
\Pi_{ij}(x)=-e2 (\partial_i \partial_j - \Box \delta_{ij}) H(x) \ \ .
\eqn
The function $H(x)$ fulfills the differential equations
\bqna
H{\prime\prime} (x)-\frac{H{\prime} (x)}{x}&=&4{\Delta{\prime}}2 (x)
\nonumber \\[1ex]
\frac{H{\prime} (x)}{x}&=&2{\Delta{\prime}}2(x)
\eqna
The compatible solution is given by
\bqn
H(x)=-\frac{1}{2\pi2}\ln (bx)
\eqn
The momentum version of Eq. (31) is obtained by Fourier transforming $H(x)$
\bqn
H(p)=\frac{1}{\pi p2}
\eqn
and integrating by parts the differential operator, thus yielding
\bqn
\Pi_{ij}(p)=\frac{e2}{\pi}\left( \frac{p_i p_j}{p2} -\delta_{ij} \right)
\eqn
{}From the full boson propagator at one loop order one gets the following
induced photon mass
\bqn
m2_\gamma=\frac{e2}{\pi} \ \ ,
\eqn
which coincides with the exact result in Ref. [8].

It has been shown that the method of differential renormalization can be
successfully applied to
theories  with topological and dynamical symmetry breaking effects.
The lower dimensional cases are particularly
instructive, specially the three dimensional theory, where the massive
propagator has a very simple form in terms of elementary functions. The
manipulations are performed in a very transparent manner, with all
expressions being finite and gauge invariant.
The coordinate space, which has been neglected for many years, is found to
be a powerful computational tool and deserves further consideration.
\vspace{1cm}

%ACKNOWLEDGEMENTS.
It is a pleasure to thank D.Z. Freedman for valuable comments and advice.
Discussions with P. Haagensen, J.I. Latorre and X. Vilasis are gratefully
acknowledeged.
E.W.N. Glover helped in the preparation of this note. This work has been
supported by CAICYT grant AEN 90-0033. A Fleming fellowship from the Spanish
Ministry of Education jointly with the British Council is also acknowledged.

\end{document}